\documentclass[aps,prx,
reprint,
showpacs,twocolumn,superscriptaddress,
amsfonts,amssymb, %eqsecnum
]{revtex4-2}
\usepackage{xr}
\usepackage{amsmath}
\usepackage{amsfonts,amssymb,amsthm,amsxtra}
\usepackage{algorithm}
\usepackage{algorithmic}
\usepackage[]{graphicx}
\usepackage{grffile}
\usepackage{mathrsfs}
\usepackage{framed}
\usepackage{bbm}
\usepackage{bm}
\usepackage{braket}
\usepackage{xcolor}
\usepackage{mathtools}
\usepackage[bookmarks=false,colorlinks=true,urlcolor=blue,citecolor=blue,linkcolor=blue]{hyperref}
\begin{document}

\title{Multi-reference Quantum Davidson Algorithm for Quantum Dynamics}

\author{Noah Berthusen}
% \email{nfbert@umd.edu}
\affiliation{Joint Center for Quantum Information and Computer Science, NIST/University of Maryland, College Park, Maryland 20742, USA}
\author{Faisal Alam}
\affiliation{Department of Physics, University of Illinois at Urbana-Champaign, IL 61801, USA}
\author{Yu Zhang}
\email{zhy@lanl.gov}
\affiliation{Theoretical Division, Los Alamos National Laboratory, Los Alamos, NM 87545, USA}

\date{\today}

\begin{abstract}
Simulating quantum systems is one of the most promising tasks where quantum computing can potentially outperform classical computing. However, the robustness needed for reliable simulations of medium to large systems is beyond the reach of existing quantum devices. To address this, Quantum Krylov Subspace (QKS) methods have been developed, enhancing the ability to perform accelerated simulations on noisy intermediate-scale quantum computers. In this study, we introduce and evaluate two QKS methods derived from the QDavidson algorithm, a novel approach for determining the ground and excited states of many-body systems. Unlike other QKS methods that pre-generate the Krylov subspace through real- or imaginary-time evolution, QDavidson iteratively adds basis vectors into the Krylov subspace. This iterative process enables faster convergence with fewer iterations and necessitates shallower circuit depths, marking a significant advancement in the field of quantum simulation.
\end{abstract}

\maketitle

\section{Introduction}
The time evolution of quantum systems is a central problem in condensed matter physics and quantum chemistry \cite{app1, app2, Ollitrault:2021un, Nelson:2020tc}. In particular, the exploration of strongly correlated quantum systems driven out of equilibrium is essential for uncovering new physics~\cite{Aoki:2014wo, Polkovnikov:2011wd, Qin_AnnRev:2022}, which has garnered significant research interest in recent days. This surge in interest requires efficient numerical modeling capabilities to understand and harness the rich physical phenomena inherent in these systems, including non-equilibrium topological phases~\cite{Rudner:2020tq, Liu:2018vc}, time crystalline order~\cite{Zaletel:2023qt}, and high harmonic generations~\cite{Gorlach:2020vi}. However, the intrinsic complexity of these problems, characterized by strong correlations and non-equilibrium, makes the simulations intractable on classical computers because representations of quantum states and operators grow exponentially with system size. 

Quantum computers were conceived to efficiently and precisely solve such problems. It is known that the problem of quantum dynamics belongs in the complexity class, Bounded-error Quantum Polynomial time (BQP)~\cite{childs}, meaning it can be solved on a quantum computer with a circuit depth that scales polynomially with system size and evolution time, promising a significant quantum advantage in efficiency over classical computing methods. However, near-term quantum computers have limited coherence times and suffer from gate infidelities, severely limiting the depth of the circuit they can execute reliably. This restricts the application of quantum algorithms for dynamics to short evolution times~\cite{Miessen:2023ul}. While foundational in the realm of quantum simulation, the conventional Trotter methods encounter significant limitations when applied to the complex dynamics of driven quantum systems. These methods, hinging on the Trotter-Suzuki decomposition~\cite{Suzuki:1976vp}, discretize the continuous evolution of a quantum system into a series of small, manageable steps. However, this approach inherently assumes a slowly varying Hamiltonian, which does not align with the rapidly changing dynamics of driven systems. In cases where the Hamiltonian varies significantly over time, as is common in driven systems, the Trotter approximation loses its fidelity, leading to large cumulative errors.

In recent years, a suite of algorithms has emerged under the umbrella of \textit{fast forwarding} \cite{cirstoiu2020, Cortes:2022uu,gibbs2021longtime, Atia:2017tt} that attempt to overcome this limitation and perform evolution to arbitrary times with constant depth circuits. While these algorithms are not universally applicable to generic Hamiltonians \cite{no_fast_forwarding}, fast forwarding allows us to effectively exploit currently available quantum resources for many relevant systems of interest.
In general, fast forwarding refers to approximating the evolution operator for a short time $t$ as~\cite{cirstoiu2020}: 
\begin{equation}
    \label{eq:fast_forwarding}
        e^{-iHt} \approx WDW^{\dagger}, 
\end{equation}
where $D$ and $W$ are quantum circuits of polynomial depth, with $D$ consisting of mutually commuting gates. This condition ensures that $D^r$ has the same circuit depth as $D$ and allows us to approximate the evolution operator for long times, $T=rt$ as $e^{-iHT} \approx WD^rW^{\dagger}$. 
The challenge lies in identifying the circuits $W$ and $D$ for a given Hamiltonian. Early work on fast-forwarding proposed building the circuits from parametrized gates, which would then be optimized with a quantum-classical feedback loop to satisfy Eq.~\eqref{eq:fast_forwarding}, akin to the Variational Quantum Eigensolver (VQE) technique~\cite{Peruzzo2014, Tkachenko:2021wq, Zhang:2022vw}. However, achieving good optimization requires executing a substantial number of quantum circuits. More recent efforts have explored using Krylov subspace (QKS) methods~\cite{jctc.9b01125, GarnetNP2020, travis2021benchmark, lin2021qks, Tkachenko:2024wq} to determine the $W$ and $D$ circuits that approximate the evolution operator well within a small subspace of the larger Hilbert space. 

In this paper, we propose enhancements to Krylov subspace methods by utilizing the QDavidson algorithm. Initially developed to identify the ground and excited states of quantum chemistry Hamiltonians, QDavidson enables the intelligent selection of additions to our Krylov subspace, thereby reducing quantum resource demands for fast-forwarding. A significant benefit of the Krylov subspace approach is its adaptability to open quantum dynamics, wherein the quantum system evolves under the influence of a bath. Our study further investigates this generalization, highlighting the practicality and versatility of the QDavidson-enhanced Krylov subspace methods in addressing complex quantum dynamics.

This paper is structured as follows: We start with detailed descriptions of both the closed and open quantum versions of the algorithms we aim to benchmark in Section~\ref{sec:quantum_krylov}. Section~\ref{sec:results} presents our numerical experiments. The paper concludes with discussing our findings and suggestions for future research in Section~\ref{sec:discussion}.

\section{Quantum Krylov Subspace Methods}
\label{sec:quantum_krylov}

In this section, we delve into the intricacies of the QKS methods, particularly their multi-reference counterparts, outlining their fundamental principles.
Multi-reference methods have played a pivotal role in advancing the accuracy and applicability of quantum chemistry simulations~\cite{Lischka:2018td}. Unlike single-reference methods that are limited by the description of a single electronic configuration, multi-reference methods consider multiple configurations simultaneously, thereby capturing complex electronic correlation effects essential for describing a wide range of chemical phenomena. These methods are particularly advantageous in handling systems with strong electron correlation, open-shell species, and systems undergoing bond-breaking and forming processes. We briefly introduce various strategies for constructing the Krylov subspace, each with distinct circuit complexities, providing a comprehensive understanding of their respective advantages, limitations, and implications for computational efficiency and accuracy. Following this, we discuss the application of the multi-reference Quantum Krylov subspace in fast-forwarding the quantum dynamics of both closed and open systems.

\subsection{Multi-reference quantum Krylov subspace methods}
Within the QKS framework, the low-lying states of a system are represented as a linear combination within the Krylov subspace. This is mathematically expressed as,
\begin{equation}
    \label{eq:krylov_subspace}
    \ket{\psi(t)} \approx \ket{\psi_K(t)} = \sum^{M}_{m=1}\sum^R_{r=1} \bm{c}_{mr}(t) \ket{\chi_{mr}}
\end{equation}
where $\bm{c}(t) \in \mathbb{C}^{RM}$ are the corresponding coefficients associated with each Krylov subspace $\{\ket{\chi_{mr}}\}$. The order-$M$ Krylov subspace for a single reference state is generated using real-time evolution, 
\begin{equation}
    \label{eq:real_time_krylov}
\mathcal{K}_M = \text{span}\{\ket{r}, e^{-iH\tau}\ket{r}, ..., e^{-iH(M-1)\tau}\ket{r}\}.
\end{equation}

In practice, we must approximate $e^{-iH\tau}$ using a Trotter decomposition of the exact time evolution. Additional reference states are iteratively added to the Krylov subspace until some convergence criteria are met; this could be when some physical observable or dynamical property is estimated to be within some tolerance, $\epsilon$. In Ref.~\cite{Cortes:2022uu}, the authors propose a simple selection process that samples the time-evolved reference state $\ket{\chi_{M-1, R-1}} = e^{-iH(M-1)\tau}\ket{r_{R-1}}$ by performing measurements in the computational basis on the quantum computer. The measurement outcomes will follow the probability distribution $p(x) = | \bra{x} e^{-iH(M-1)\tau}\ket{r_{R-1}} |^2$. The bitstring with the largest observed probability has its order-$M$ Krylov subspace, Eq.~\eqref{eq:real_time_krylov}, added to the full Krylov subspace.

\subsection{Quantum Davidson Algorithm}
Unlike the previous method, the Krylov subspace is not pre-generated using time evolution. Instead, the QDavidson algorithm iteratively grows the Krylov subspace in such a way that it stays close to the exact eigenspace. Doing this has the advantage of faster convergence and fewer reference states required at the cost of more circuit evaluations. Previously, the QDavidson algorithm has been used to express the ground and excited states of the Hamiltonian $H$ as a linear combination of reference states in the Krylov subspace \cite{Tkachenko:2024wq},
\begin{equation}
    \ket{\psi_i} = \sum_{k=1}^M \bm{v}_{ki} \ket{\chi_k}.
\end{equation}
in much the same way as the original single-reference Quantum Davidson algorithm~\cite{Tkachenko:2024wq}. 
The intuition is that if these states are well expressed, then time evolutions (within the eigenspace found by the algorithm) will be similarly precise.
An iteration of the QDavidson algorithm begins by computing on the quantum computer the subspace matrices $D$ and $E$, whose matrix elements are defined as:
\begin{equation}
    \label{eq:dande}
    D_{k\ell} = \bra{\chi_k} H \ket{\chi_\ell} \ \text{and} \ E_{k\ell} = \braket{\chi_k | \chi_\ell}.
\end{equation}
This can be done using repeated SWAP tests~\cite{Buhrman_2001}, Hadamard tests~\cite{Anaronov_2006}, or the MFE protocol~\cite{Cortes:2022uu}. We now solve the generalized eigenvalue problem $D\bm{v} = \lambda E \bm{v}$ to get the approximate eigenstates and associated energies. Afterward, the residue of each of the computed eigenstates can be evaluated,
\begin{equation}
    \ket{R_i} = H\ket{\psi_i} - \lambda_i\ket{\psi_i} = (H - \lambda_i)\sum_{k=1}^M \bm{v}_{ik} \ket{\chi_k}.
\end{equation}
The norm of the residue is then measured on the quantum computer in a similar manner to the $D$ matrix,
\begin{align}
    |\ket{R_i}| &= \braket{\bm{v_i} | (H-\lambda_i)^2 | \bm{v_i}} \\
                &= \sum_{k,\ell=1}^M \bm{v}_{ki}^* \bm{v}_{\ell i} \braket{\chi_k | (H - \lambda_i)^2 | \chi_\ell} 
\end{align}
Consequently, the correction vector ($\ket{\delta_i}$) and the associated pre-conditioned correction vector ($\ket{\delta'_i}$) can be derived as
\begin{align}
   \label{eq:residual}  \ket{\delta_i} &= e^{-\Delta\tau (\hat{H}-E_I)}\ket{\Psi_I},  
    \\ 
    \ket{\delta_i'}& =  \ket{\delta_{I}}-\sum_{KJ} \ket{\Psi_K}(S^{-1})_{KJ}\bra{\Psi_J}\delta_{I}\rangle.
\end{align}
If $|\ket{\delta'}| < \epsilon$, then the correction vector $\ket{\delta}$ cannot be expressed as a linear combination of the reference states in the current Krylov subspace. As such, $\ket{\delta}$ is then added as a new reference state after mapping the non-unitary operator in Eq.~\eqref{eq:residual} into unitaries~\cite{GarnetNP2020}. Doing this for each of the current reference states completes a full iteration of the algorithm. Iterations are continued until the resulting simulation is sufficiently accurate for some desired physical observable and target final time.  

With the subspace now constructed, we can estimate the time dynamics of the system by expressing it as a linear combination of reference states, as we did in Eq.~\eqref{eq:krylov_subspace}. However, in this case, the coefficients are the expansion of the Wavefunction in the eigenstate computed from the Quantum Davidson algorithm,
\begin{equation}
    \label{eq:davidson_subspace}
    \ket{\psi(t)} \approx \ket{\psi_K(t)} = \sum^M_{i=1} \bm{c}_i(t) \ket{\chi_i} .
\end{equation}

\subsection{Multi-reference Quantum Davidson method}

A natural extension of the QDavidson algorithm involves integrating with the multi-reference Krylov method, incorporating time-evolved reference states into the subspace. After a certain number of iterations, QDavidson yields a progressively expanded set of reference states $\{\ket{\chi_i}\}$. Each reference state then forms the basis for constructing the order-$M$ Krylov subspace, following the process described in Eq.~\eqref{eq:real_time_krylov}. 
The accuracy of the fast-forwarding process is evaluated to determine if it meets the desired level of precision. If the accuracy is lacking, an additional QDavidson iteration is undertaken, utilizing the existing reference states $\{\ket{\chi_i}\}$. Compared to the conventional multi-reference Krylov approach, the distinct aspect of this method lies in the selection process of the reference states.
By adopting this strategy, we aim to leverage the advantageous features of both methodologies, thus enhancing the efficacy and precision of the fast-forwarding process in quantum dynamics simulations.

\subsection{Fast Forwarding for closed-system dynamics}

Once the Krylov subspace is constructed using any of the mentioned algorithms, the subsequent steps in the fast-forwarding process remain the same. With the subspace matrices \(D\) and \(E\), one can address a generalized eigenvalue problem to represent the matrix exponential \(e^{-iHt}\) in the eigenbasis of the Krylov subspace~\cite{Lim_2021}. Alternatively, the quantum subspace Schrödinger equation can be solved,
\begin{equation}
    i S \partial_t \bm{c}(t) = D \bm{c}(t),
\end{equation}
which has the following analytical solution:
\begin{equation}
    \label{eq:subspace_schro}
    \bm{c}(t) = e^{-iS^{-1}Dt}\bm{c}(0).
\end{equation}
where $\bm{c}(0)$ are the initial expansion coefficients. When the initial state is one of the Krylov basis states $\ket{\chi_i}$, then $\bm{c}(0)$ is the all zeros vector with a one in the $i$th position. Generally, the subspace overlap matrix $S$ can be poorly conditioned and might not have a well-defined inverse. To remedy this, we use the singular value decomposition of $E$ and zero out small singular values before taking the inverse (see \cite{klymko2021real, Cortes:2022uu_2}). Then, to fast-forward the system to arbitrary times $t$, we apply the solution of the subspace Schrödinger equation Eq.~\eqref{eq:subspace_schro} to the expansion coefficients themselves. This yields a time-evolution of the coefficients $\bm{c}(t)$ which can be used to express an estimate of $\ket{\psi(t)} = e^{-iHt}\ket{\psi(0)}$ as a linear combination of the reference states in the Krylov subspace as noted in Eq.~\eqref{eq:davidson_subspace}.

One of the main motivations for performing time dynamics of quantum systems is to calculate time-dependent observables. In other fast-forwarding methods where the time-evolved state is available~\cite{cirstoiu2020, gibbs2021longtime}, the observables of interest can be computed directly. However, this is not true for QKS methods since generating the corresponding quantum state from the time-evolved expansion coefficients and reference states is not straightforward. Nevertheless, a general time-evolved observable $O(t)$ can be expressed as follows:
\begin{align}
    O(t) &= \braket{\psi_K(t)|O|\psi_K(t)} \nonumber \\
    \label{eq:observable}
         &\approx \sum_{ij} \bm{c}_i^*(t) \bm{c}_j(t) \braket{\chi_i|O|\chi_j}.
\end{align}
We have used Eq.~\eqref{eq:davidson_subspace} for simplicity, but this technique works for any of the QKS methods, multi-reference or not. The values of $\braket{\chi_i|O|\chi_j}$ can be calculated on the quantum computer in the same way as the subspace matrix $D$, and then the results are post-processed using the time-evolved coefficients to give the observable $O(t)$.

A benefit of using QKS methods for fast-forwarding is that once the $D$ and $E$ matrices and Eq.~\eqref{eq:observable} are calculated, only classical post-processing is required to determine $O(t)$ for any time $t$ less than the target final time. Computing the time-evolved coefficients, Eq.~\eqref{eq:subspace_schro}, is done entirely on the classical computer. If a different observable is of interest, it is only necessary to recompute Eq.~\eqref{eq:observable} on the quantum computer. This is an advantage over other fast-forwarding methods, where a new time-evolved state must be prepared for each observable and each time.

\subsection{Quantum dynamics of open quantum system}
This subsection will briefly discuss how the QKS method (particularly Quantum Davidson) for quantum dynamics developed above can be applied to open quantum systems within Lindblad and generalized quantum master equation (GQME) schemes. The main challenge in simulating open quantum system dynamics on quantum computers is managing the non-Hermitian time propagation~\cite{Wang:2023wf, Hu:2020vx, Schlimgen:2021vs, Cattaneo:2023wv, Del-Re:2024wn}, as the fundamental operations (gates) of quantum computers are inherently unitary. To address this challenge, we can leverage the techniques developed for mapping non-unitary operators into unitary ones. Specifically, here, we extend the mapping of non-unitary operators into unitaries, as described in Eq~\eqref{eq:residual}, to the non-Hermitian time propagation to simulate the dynamics of open quantum systems.

The standard approach for deriving the quantum dynamics of an open quantum system interacting with its environment starts from the von Neumann equation of the combined system,
\begin{equation}\label{eq:qme_all}
    i\hbar\partial\rho_{\text{tot}}(t)=[\hat{H}_{\text{tot}},\rho_{\text{tot}}]\equiv \mathcal{L}\rho_{\text{tot}}(t),
\end{equation}
where $\hat{H}_\text{tot}$ is the total Hamiltonian of the combined system, $\hat{H}_\text{tot}=\hat{H}_{S}+\hat{H}_B+\hat{H}_{SB}$, that includes the Hamiltonians of the system $\hat{H}_S$, bath $\hat{H}_B$, and system-bath interaction $\hat{H}_{SB}$. $\rho_{\text{tot}}$ is the density matrix of the entire system ($S+B$). In general, the system-bath interaction can be written as $\hat{H}_{SB}=\sum_{k\alpha} g_{k\alpha}\hat{A}_k \hat{B}_\alpha$,
where $\hat{A}_k$ are the system operators which the bath operators $\hat{B}_\alpha$ couple to, and $g_{k\alpha}$ is the corresponding coupling strength. Since we are only interested in the dynamics of the system, we can perform a partial trace over the bath degrees of freedom in Eq.~\eqref{eq:qme_all} and thereby obtain a master equation for the motion of the original system density matrix. 

The non-Markovian dynamics of an open quantum system can be described by the generalized quantum master equation (GQME) (Nakajima-Zwanzig)~\cite{Breuer2007, Nakajima:1958to, Zwanzig:1960vw}. We define a projector operator $\mathcal{P}$ that projects the total density matrix into the system subspace.
The complementary projection superoperator is denoted as $\mathcal{Q}=1-\mathcal{P}$. $\mathcal{P}$ and $\mathcal{Q}$ satisfy
$\mathcal{P}^2=\mathcal{P}, \mathcal{Q}^2=\mathcal{Q}$, and $[\mathcal{P}, \mathcal{Q}]=0$.
Hence, projecting Eq.~\eqref{eq:qme_all} into the subsystem and following the standard derivation within the Nakajima-Zwanzig formalism~\cite{Breuer2007, Nakajima:1958to, Zwanzig:1960vw}, we can get the dynamics of the reduced density matrix (for the system only) as
\begin{equation}\label{eq:gqme}
    i\hbar\partial\rho_{\text{S}}(t)=\mathcal{L}\rho_{\text{S}}(t)
    =\mathcal{L}_{0}\rho_{\text{S}}(t) + \mathcal{L}_{m}[\rho_{\text{S}}(t)],
\end{equation}
where $\mathcal{L}$ is the Liouvillian superoperator, including the overall system Liouvillian $\mathcal{L}_0=\text{Tr}_{B}[\hat{H},\cdot]$ (time-independent) and memory kernel (time-dependent) $\mathcal{L}_{m}$ parts. Here, $\text{Tr}_{B}$ denotes the partial trace over the bath degrees of freedom. The memory kernel, in general, can be written as
\begin{equation}
    \mathcal{L}_{m}[\rho_{\text{S}}(t)]=-i\hbar\int^t_0\mathcal{K}(\tau)\rho_{\text{S}}(t-\tau)d\tau.
\end{equation}
The memory kernel is given by
\begin{equation}
    \mathcal{K}(\tau)=\frac{1}{\hbar^2}\text{Tr}_B\left\{
    \mathcal{L}e^{-i\mathcal{Q}\mathcal{L}\tau/\hbar}\mathcal{Q}\mathcal{L}\rho_B(0)
    \right\},
\end{equation}
where $\mathcal{Q}$ is the projection operator on the bath degrees of freedom, and $\mathcal{L}$ is the overall system-bath Liouvillian (note: the overall system-bath Liouvillian is Hermitian).

When memory effect is not essential, we can further simpilfy the quantum dynamics with Markovian approximation. The most general trace-preserving and completely positive form of Markovian evolution is the Lindblad master equation for the reduced density matrix $\rho_{\text{S}}=\text{Tr}_B[\rho_\text{tot}]$,
\begin{align}\label{eq:lindblad}
    i\hbar\dot{\rho}_{\text{S}}(t)=& \mathcal{L}_0\rho_{\text{S}}(t)
    \nonumber\\
    &+i\hbar\sum_k \left(C_k\rho_{\text{S}}(t)C^\dag_k - \frac{1}{2}\{C^\dag_k C_k,\rho_{\text{S}}(t)\}\right).
    % \mathcal{L}_{diss}\rho(t)
\end{align}
where $C_k=\sqrt{\gamma_k}\hat{A}_k$ are the collapse operators and $\hat{A}_k$ are the operators through which the bath couples to the system in $\hat{H}_{int}$.

Nevertheless, no matter whether it is in Lindblad or GQME formalism, the resulting time-dependent density matrix can be written in the following generalized form with a non-unitary propagator, i.e.,
\begin{equation}
    \rho_{\text{S}}(t)=e^{-i/\hbar\mathcal{L}t}\rho_{\text{S}}(0)
    =e^{-i/\hbar(\mathcal{L}_{1}+\mathcal{L}_2)t}\rho_{\text{S}}(0).
\end{equation}
where $\mathcal{L}_1$ and $\mathcal{L}_2$ describe the unitary and non-unitary evolution operators, respectively. It should be noted that the Lindblad master equation is identically the vectorization mapping in the Liouville space, resulting in
\begin{align}\label{eq:vectorrho}
    \frac{d\ket{\rho_{\text{S}}}}{dt}=&\Big[
    -i I\otimes \hat{H} + i\hat{H}^\intercal\otimes I + \sum_k\Big(\overline{L_k}\otimes L_k
    \nonumber\\
    & -\frac{1}{2}I\otimes (L^\dag_k L_k) - \frac{1}{2}(L^\intercal_k\overline{L_k})\otimes I \Big)\Big]\ket{\rho_{\text{S}}}.
\end{align}
where the bar indicates entrywise complex conjugation.

However, this requires a doubling of the qubits and an overhead of an ancilla and controlled operators for evaluating observables. Alternatively, we can map the density matrix operator as (isomorphism mapping)
\begin{equation}
    \rho_{\text{S}}=\sum_{p} C_p U\ket{p}\bra{p}U^\dag
    \rightarrow \quad
    \ket{\rho_{\text{S}}}=\sum_{p} C_p U\ket{p}\otimes \bar{U}\ket{p},
\end{equation}
where $\ket{p}$ are the $n$-qubit computational basis states in the $2^n$ possible bit strings. The corresponding Lindblad equation then becomes Eq.~\eqref{eq:vectorrho}.

After the isomorphism mapping, the propagator can be again Trotterized, and each term can be applied term by term.
\begin{align}
   & e^{-i(I\otimes\hat{H} - i\hat{H}^{\intercal}\otimes I)\tau}\sum_p C_p U\ket{p}\otimes \bar{U}\ket{p}
   \nonumber\\
   &=\sum_p C_p e^{i\hat{H}^{\intercal} \tau}U\ket{p} \otimes e^{-i\hat{H}\tau}\bar{U}\ket{p}.
\end{align}
Where $e^{i\hat{H}^\intercal\tau}=\overline{{e^{-i\hat{H}\tau}}}$ for Hermitian $\hat{H}$. The remaining Lindblad terms in the Trotterized propagator are of the form $e^{-L^{\intercal}_k L_k\tau/2}\otimes e^{-L^\dag_k L_k\tau/2}$ and $e^{\bar{L}_k\otimes L_k\tau}$. The first term preserves the ansatz but is non-unitary, while the second term does not preserve the ansatz and is non-unitary. The QITE algorithm can then be used to map the non-unitary terms into unitaries. Suppose the non-unitary term is denoted as $V_k$. For example, $V_k=e^{-L^{\intercal}_k L_k\tau/2}\otimes e^{-L^\dag_k L_k\tau/2}$, we can use the QITE algorithm to find a set of numbers $D_p$ and a Hermitian operator $A$ such that
\begin{align}
    V_k \sum_p C_p U\ket{p}\otimes \bar{U}\ket{p} =& \sum_p (D_p + C_p)e^{iA}U\ket{p}
    \nonumber\\
    &\otimes e^{-i\bar{A}}\overline{U}\ket{p} + \mathcal{O}(\tau^2).
\end{align}
where
\begin{equation}
    D_q = -\tau C_p \text{Re}\left[\bra{p}U^\dag L^\intercal_k\overline{L_k}U\ket{p}\right].
\end{equation}
Like the QITE method, the $A$ operator can be solved using a linear algebra~\cite{Kamakari:2022vg}.

\section{Results}
\label{sec:results}

\begin{figure}[!t]
    \centering
    \includegraphics[width=0.95\linewidth]{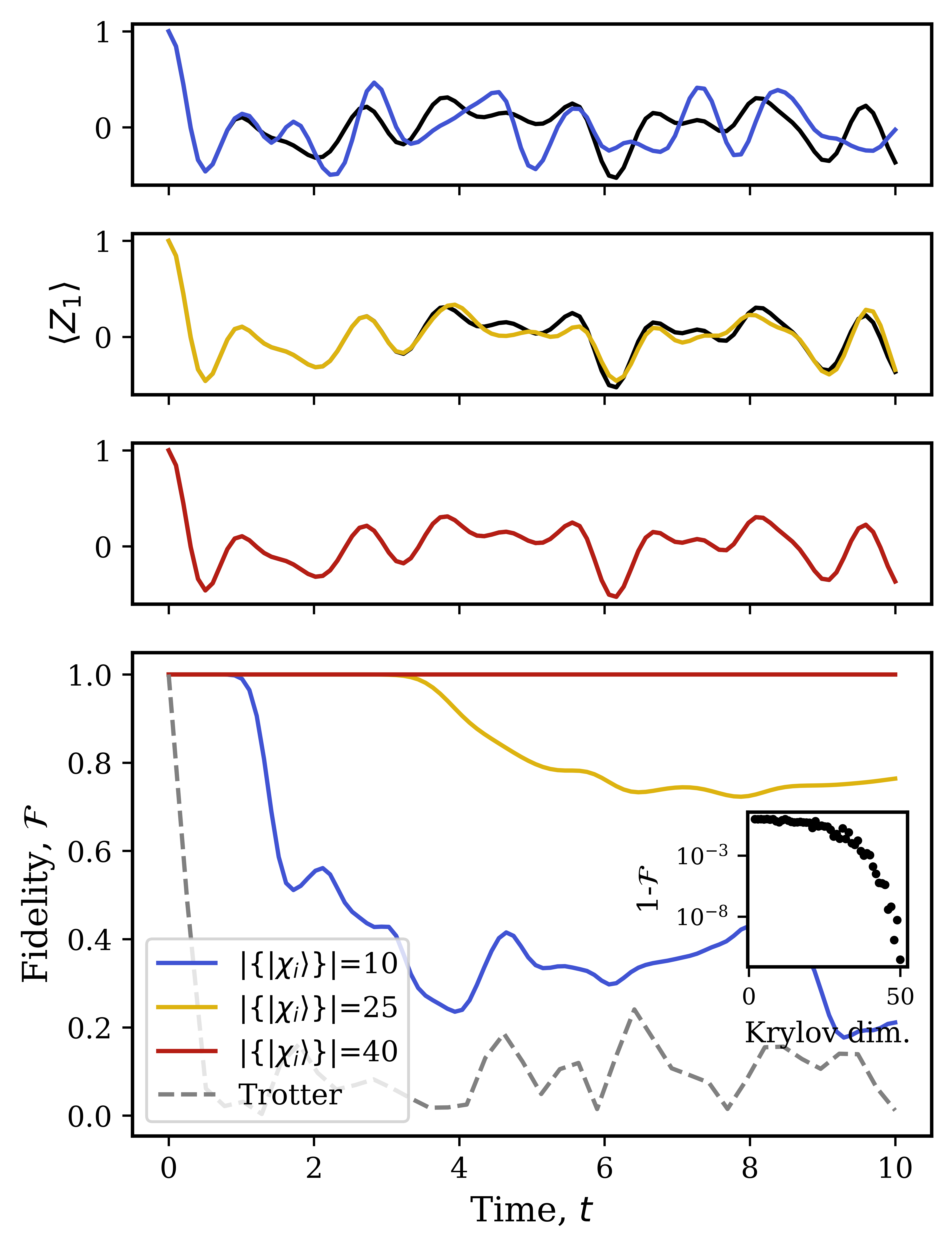}
    \caption{Results of performing fast-forwarding of the XYZ model on $N=8$ qubits, setting $J_X = J_Y = J_Z = h = 1$, and using a Krylov subspace generated by the QDavidson algorithm after running for several iterations. We compare subspaces of dimension 10 (blue), 25 (yellow), and 40 (red) against an exact time-evolved state (black). The top three panels show the expectation value $\langle Z_1 \rangle$ for each state. The bottom panel indicates the state overlap between the exact time-evolved state with the fast-forwarded states, $\mathcal{F} = |\braket{\psi(t) | \psi_K(t)}|^2$. The gray dashed line indicates a state overlap with a time evolution using a first-order Trotter approximation and a comparable number of Trotter steps. The inset is a semilog plot of infidelity at time $t_f = 10$ as a function of the Krylov subspace dimension. }
    \label{fig:qdavidson_sims}
\end{figure}

We first benchmark the QDavidson algorithm by simulating the time evolution of the Heisenberg model,
\begin{equation}
    H = \sum_{i=1}^{N-1} J_x X_i X_{i+1} + J_y Y_i Y_{i+1} + J_z Z_i Z_{i+1} + \sum_{i=1}^N hZ_i,
\end{equation}
where $X_i, Y_i, Z_i$ are Pauli matrices acting on qubit $i$, $J_X, J_Y, J_Z, h$ are coefficients, and $N$ is the number of qubits. We initialize the algorithm in the N\'{e}el state $\ket{\psi_i} = \ket{0101\cdots}$ and simulate to a final time $t_f = 10$. In this section, all computations were done on the classical computer, including building the $D$ and $E$ matrices, Eq.~\eqref{eq:dande}, and computing the residues. To verify the accuracy of the fast-forwarded state obtained by Eq.~\eqref{eq:davidson_subspace}, we compute an exact time evolution of the system, $\ket{\psi(t)} = e^{-iHt} \ket{\psi_i}$.

In Fig.~\ref{fig:qdavidson_sims}, we present the results of fast-forwarding the Heisenberg model on $N=8$ qubits while varying the number of iterations (the dimension of the Krylov subspace) of the QDavidson algorithm.
After performing the fast-forwarding, one can calculate time-dependent properties of interest by using Eq.~\eqref{eq:observable}.
The upper three panels show the computed expectation value $\langle Z_1 \rangle$ for each fast-forward state and a black line indicating the true value.
The bottom panel reports the fidelity of each fast-forwarded state with respect to the exact time-evolved state, $\mathcal{F} = |\braket{\psi(t) | \psi_K(t)}|^2$.
Increasing the number of iterations predictably increases the accuracy of the fast-forwarding, which is reflected in a more precise match with the true expectation value and higher fidelity with the true state.
The gray dashed line indicates the overlap with the exact Wavefunction of an evolution done using a first-order Trotter approximation and 40 Trotter steps. We see that the evolution done with QDavidson vastly outperforms Trotter, even when using a significantly shorter circuit.
The inset in the bottom panel shows infidelity at time $t_f = 10$ as a function of the dimension of the Krylov subspace. We observe a slow convergence rate before a sharp drop-off.
At short times, the fidelity remains high even when using a less expressive (smaller) subspace; if one is only interested in the dynamics during this regime, then using fewer iterations and a smaller Krylov subspace is acceptable.

\begin{figure}[!t]
    \centering
    \includegraphics[width=\linewidth]{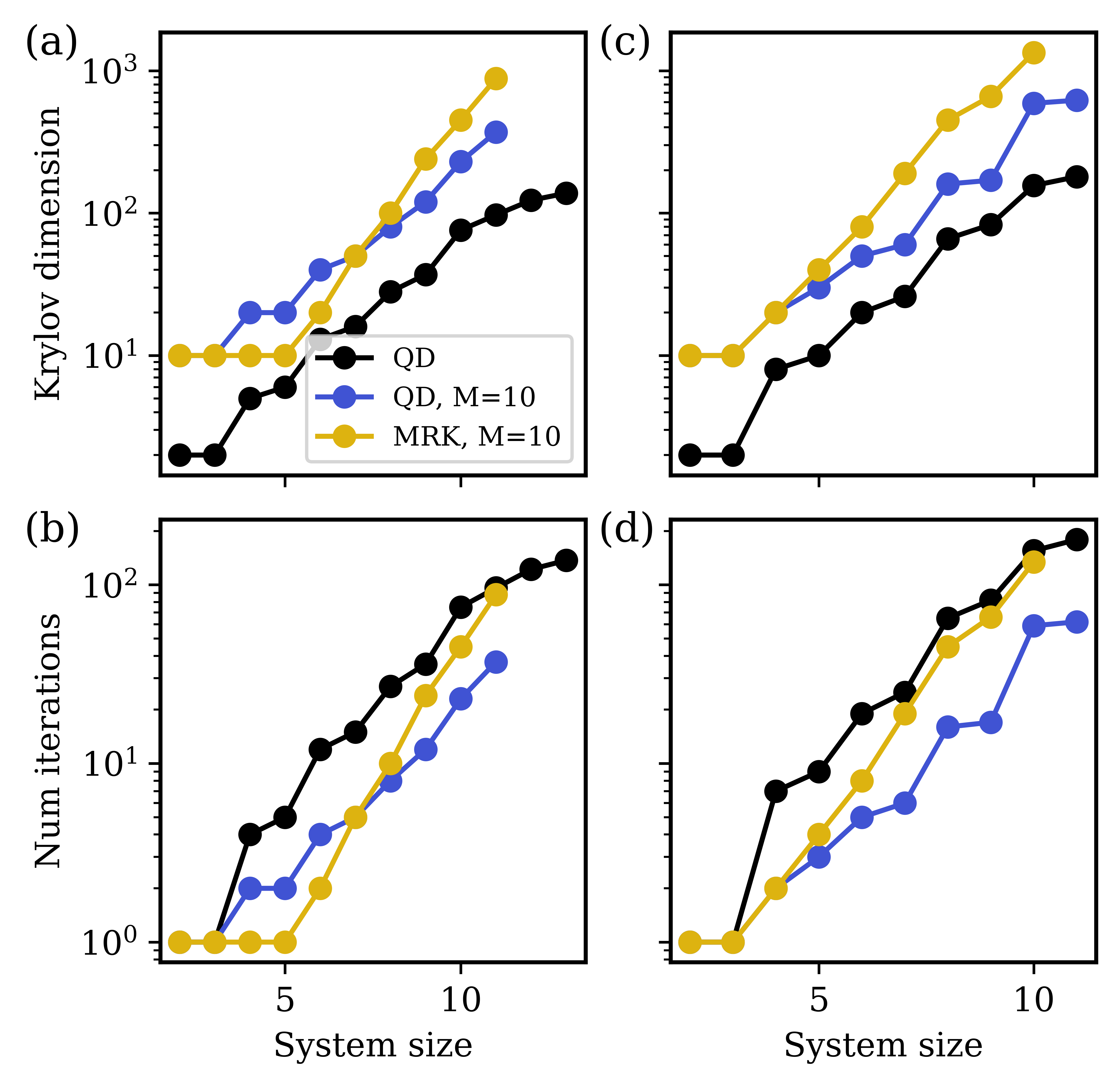}
    \vspace{-10pt}
    \caption{(a)-(b) Scaling of QKS methods with system size on the Heisenberg XXX model with $J_x = J_y = J_z = h = 1$. (a) The semilog plot required Krylov dimension to fast-forward to time $t_f=10$ with fidelity 0.9 as a function of system size. Here, we compare the QDavidson algorithm (black markers) against multi-reference QDavidson (blue markers) and multi-reference Krylov (yellow markers). Both multi-reference methods use $M=10$ reference states. (b) The semilog plot of a corresponding number of algorithm iterations to reach the desired fidelity. (c)-(d) Similar scaling of QKS methods with system size on the Heisenberg XYZ model with $J_x = h = 1, J_y = 2, J_z = 3$. }
    \label{fig:qd_vs_mrk}
\end{figure}

To compare the scaling of (multi-reference) QDavidson and multi-reference Krylov for time dynamics, we simulate systems of up to $N=13$ qubits.
We focus here on the Heisenberg model where the Hamiltonian coefficients are set to be equal, $J_x = J_y = J_z = h = 1$.
In Fig.~\ref{fig:qd_vs_mrk}(a), we plot the minimum Krylov dimension required for the fast-forwarded state to have 90\% overlap with the exact Wavefunction at the final time $t_f = 10$. We compare the standard QDavidson algorithm (black markers) against multi-reference Krylov with $M = 10$ reference states (yellow markers) as well as multi-reference QDavidson again with $M = 10$ reference states (blue markers). For the multi-reference methods, the order-$M$ Krylov subspace was generated using an exact time evolution; we later investigated the effect of Trotter error on the methods' performance.
Regarding the number of states needed to recreate the exact time-evolved state accurately, QDavidson outperforms the other methods by almost an order of magnitude. The semilog plot clearly shows sub-linear behavior from the QDavidson algorithm, indicating sub-exponential scaling of the Krylov dimension for this Hamiltonian. The same cannot be said about the multi-reference methods, where the scaling is unclear.
Fig.~\ref{fig:qd_vs_mrk}(b) displays the corresponding number of algorithm iterations required to reach the desired 90\% overlap. Here, we see that the multi-reference methods need fewer iterations than the QDavidson algorithm, although, for large enough system sizes, this will no longer hold.  

Figs.~\ref{fig:qd_vs_mrk}(c)-(d) show similar trends for the Heisenberg XYZ model with $J_x = h = 1, J_y = 2, J_z = 3$. The scaling behavior of the required Krylov dimension and the number of iterations is consistent with Figs.~\ref{fig:qd_vs_mrk}(a)-(b). The QDavidson algorithm continues to demonstrate sub-exponential scaling in the Krylov dimension, outperforming the multi-reference methods regarding the number of states needed to represent the time-evolved state accurately. The number of iterations required for the multi-reference methods remains lower, but this difference diminishes with increasing system size. These results indicate that the QDavidson algorithm is robust and efficient across different models and parameter settings, providing a significant advantage for simulating quantum dynamics in larger systems.

\begin{figure}[htb]
    \centering
    \includegraphics[width=0.9\linewidth]{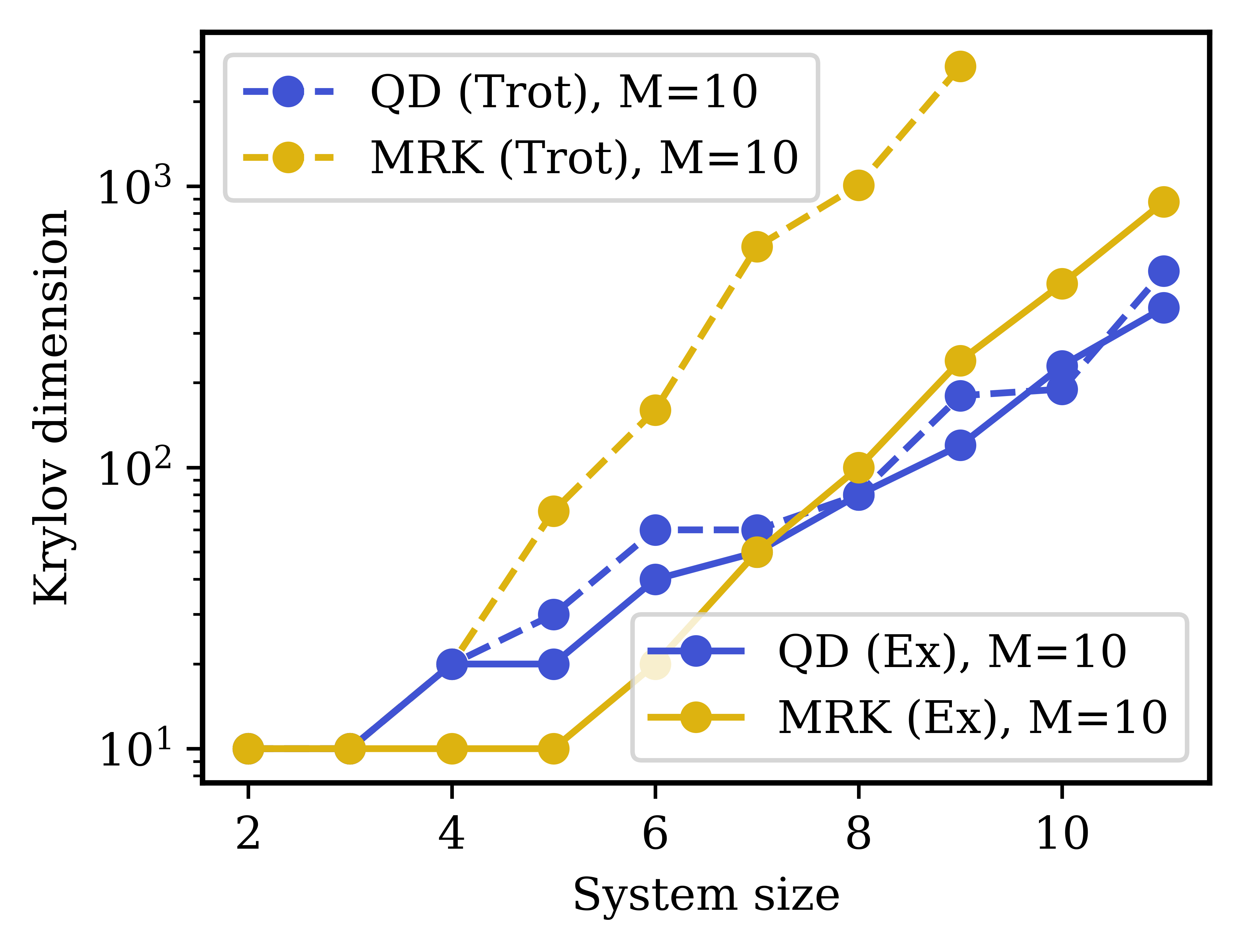}
    \vspace{-10pt}
    \caption{Semilog plot comparing the effect of Trotter error on the required Krylov dimension to fast-forward to time $t_f = 10$ with a fidelity of 0.9. Here we again use the Heisenberg XXX model with $J_x = J_y = J_z = h = 1$. We compare the multi-reference Krylov method with $M=10$ reference states where the order-$M$ Krylov subspace was generated with exact time evolution (solid yellow) against one where the Krylov subspace was generated with a first-order Trotter product formula and Trotter step size $\tau=0.1$ (dashed yellow). We also compare an exact time evolution-generated multi-reference QDavidson (solid blue) with a Trotter-generated multi-reference QDavidson (dashed blue).}
    \label{fig:trotter_comp}
\end{figure}

\subsection{Effect of Trotter error}
\label{sec:trotter}

In the previous section, we focused on the performance of the algorithms by neglecting the effects of noise and error. Here, we discuss the impact of the Trotter error on the Krylov subspace dimensions. Fig.~\ref{fig:trotter_comp} compares the required Krylov dimension to fast-forward to time $t_f = 10$ with a fidelity of 0.9 for the Heisenberg XXX model with $J_x = J_y = J_z = h = 1$. We compare the multi-reference Krylov method with $M=10$ reference states, where the order-$M$ Krylov subspace was generated with exact time evolution (solid yellow) against one where the Krylov subspace was generated with a first-order Trotter product formula and Trotter step size $\tau=0.1$ (dashed yellow). We also compare an exact time evolution-generated multi-reference QDavidson (solid blue) with a Trotter-generated multi-reference QDavidson (dashed blue). 

The results indicate that the multi-reference QDavidson algorithm is robust against Trotter errors, maintaining its efficiency and accuracy even when approximations are introduced. For this method, the required Krylov dimension is practically unchanged in the presence of Trotter errors. 
While still effective, the multi-reference methods show a more pronounced increase in the required Krylov dimension when Trotter errors are present.

\section{Discussion}
\label{sec:discussion}

In this work, we have developed and benchmarked the QDavidson algorithm for fast-forwarding quantum simulations. Compared with the recently introduced selected quantum Krylov fast-forwarding (sQKFF) algorithm, we find that the QDavidson algorithm can achieve comparable accuracies while using a smaller Krylov subspace. However, this comes at the cost of more iterations and circuit evaluations on the quantum computer. To address the trade-off between the Krylov subspace dimension and the number of iterations, we introduced a multi-reference approach that combines properties of both QDavidson and multi-reference Krylov methods. We found it to be a compromise, balancing the Krylov subspace dimension and the number of iterations required. Depending on the limitations of the available device, the number of reference states can be tuned to optimize performance.

We also examined the effect of Trotter error on the performance of these algorithms. Interestingly, we found that Trotter error can benefit the algorithm by making the resulting states less dependent. The multi-reference QDavidson algorithm, in particular, proved robust against Trotter errors, maintaining its efficiency and accuracy even when approximations are introduced. This robustness highlights the potential of QDavidson for practical implementations on near-term quantum devices.

Future work on using Krylov subspace methods would benefit from investigating the performance of these algorithms on real quantum hardware. Understanding the impact of hardware-specific noise and errors and developing error mitigation techniques will be crucial for successfully applying these methods in practical quantum simulations. Additionally, exploring the scalability of the QDavidson algorithm for larger systems and its integration with other quantum algorithms could further enhance its utility in solving complex quantum problems.

\section*{Acknowledgements}
We acknowledge support from the US DOE, Office of Science, Basic Energy Sciences, Chemical Sciences, Geosciences, and Biosciences Division under Triad National Security, LLC (``Triad") contract Grant 89233218CNA000001 (FWP: LANLECF7) and the Laboratory Directed Research and Development (LDRD) program of Los Alamos National Laboratory (LANL). This research used computational resources provided by the Institutional Computing (IC) Program and the Darwin testbed at Los Alamos National Laboratory (LANL), which is funded by the Computational Systems and Software Environments subprogram of LANL's Advanced Simulation and Computing program. LANL is operated by Triad National Security, LLC, for the National Nuclear Security Administration of the U.S. Department of Energy (Contract No. 89233218CNA000001).

\section*{Data Availability} The source code and data to generate the figures in the paper are provided freely at \url{https://github.com/noahberthusen/qdavidson-dynamics}. 
% The data supporting other findings of this study are available from the corresponding authors upon reasonable request.

\bibliography{bibliography}

\end{document}